\documentclass{eas}
\usepackage{graphicx}
\usepackage{natbib,pslatex}
\newcommand{\msini}{\ensuremath{M \sin{i}}}

\newcommand{\mearth}{\ensuremath{\mbox{M}_{\oplus}}}
\newcommand{\mjup}{\ensuremath{\mbox{M}_{\mbox{\scriptsize Jup}}}}
\newcommand\editorstart{ \bf \color{myred} } 
\newcommand\editorstop{ \rm \color{black} } 
\usepackage{color}  
\definecolor{myred}{rgb}{0.7,0.0,0.2} 
\reversemarginpar
\newcounter{exno}
\let\oldmarginpar\marginpar
\renewcommand\marginpar[1]{\addtocounter{exno}{1}%
\-\oldmarginpar[\raggedleft #1]%
{\raggedright {\sf \color{blue} {\bf \Roman{exno}}. #1}}}
\begin{document}
%
\title{A Survey of Multiple Planet Systems} 
\author{Jason T. Wright}%
\address{525 Davey Lab, Astronomy and Astrophysics, Eberly College of
  Science, The Pennsylvania State University, University Park, PA 16802
\email{jtwright@astro.psu.edu}}
%
\editorstart
\runningtitle{A Survey of Multiple Planet Systems}
\editorstop
\begin{abstract}

To date, over 30 multiple exoplanet systems are known, and 28\% of
stars with planets show significant evidence of a second companion.  I briefly
review these 30 systems individually, broadly grouping them into five
categories:  1) systems with 3 or more giant ($\msini > 0.2 \mjup$)
planets, 2) systems with two giant planets in mean motion resonance
(MMR), 3) systems with two giant planets not in MMR but whose dynamical evolution 
is affected by planet-planet interactions, 4) highly hierarchical
systems, having two giant planets with very large period ratios ($ >
30:1$), and 5) systems of ``Super-Earths'', containing only planets
with ($\msini < 20 \mearth$).    

It now appears that eccentricities are not markedly higher
among planets in known multiple planet systems, and that
planets with $\msini < 1 \mjup$ have lower eccentricities
than more massive planets.  The distribution of semimajor axes for
planets in multi-planet systems does not show the 3-day pile-up or the
1 AU ``jump'' of the apparently-single planet distribution.
\end{abstract}
\maketitle
\section{Introduction}

The first multiple exoplanet system was discovered in 1992, when
\cite{Wolszczan92} detected two very low-mass objects orbiting the
pulsar PSR B1257+12 using pulse timing methods.  Seven years later,
\cite{Butler99} announced the first multi-exoplanet system
around a normal star from radial velocities, $\upsilon$ Andromedae.  

Since then, the ever-improving precision and ever-growing temporal
baseline of radial velocity searches has rapidly increased the number of known
multiple planet systems, and today at least\footnote{This and the other numbers herein will be out of date at the time of this manuscript's publication -- indeed, many
  were out of date by the end of the conference!  At least three
  multiple planet systems announced shortly after the conference
  require mention, though they are not included in the statistics
  here:  Bouchy et al.\ announced that HD 47186 and HD 181433 harbor
  highly hierarchical systems, and \cite{Marois08} imaged three young
  objects orbiting the A  star HR 8799.} 30 such systems
are known\footnote{We
  should not neglect a 31st known system, and the best understood -- our
  own Solar System, whose planets' circular orbits belie a violent
  history of planetary collisions and ejections.}.  Thus, this conference comes at a special time in the field
of multiple planet systems:  there are just few enough systems now
that this is likely the last conference where each system can still
plausibly be discussed individually, but there are enough such that
this is the first conference where we can construct something like a
statistically significant sample of planets in multiple systems to
divide into subsamples or to compare with apparently single-planet systems.

Multiple planet systems provide an increasingly powerful way to
probe the dynamical origins of planets (e.g. \cite*{Ford06}).
Single-planet systems, because their orbits are strictly periodic,
provide most of their information on the typical migration and
interaction histories of exoplanets statistically as an ensemble.  But
each individual multiple planet system has the potential to serve as a
case study of planetary system evolution. 

\section{The Known Multi-Planet Systems}

I have divided the known multi-planet systems heuristically into
five broad, nonexclusive categories:  1) systems with 3 or more
giant ($\msini > 0.2 \mjup$) planets, 2) systems with two giant
planets in mean motion resonance,  3) systems with two giant planets
not in MMR, but whose dynamical evolution is affected by weaker planet-planet
interactions, 4) highly hierarchical systems, having two giant planets
with very large period ratios ($ > 30:1$), and 5) systems of
``Super-Earths'', containing only planets with ($\msini < 20
\mearth$).  The summaries below represent a brief and necessarily
incomplete description of each system. 

\subsection{3+ Giant Planets}

Five systems are known to comprise three or more giant planets.  

\subsubsection{$\upsilon$ Andromedae}

\cite{Butler99} announced the first multi-exoplanet system
around a normal star, $\upsilon$ Andromedae, detected using precise radial
velocity measurements from Lick Observatory.  The pattern of the discovery is
typical 
of many multiplanet systems:  after a strong, short-period signal from
a 4.6 d, $\msini = 0.68 \mjup$ planet was discovered
(\cite{Butler97}), continued monitoring revealed significant structure
in residuals that far exceeded the expected measurement uncertainties.
After 2.5 y, two additional, superimposed Keplerians consistent with
planets of $\sim 2 \mjup$ and $\sim 4 \mjup$ became apparent with
periods of $\sim 240$ d and $\sim 1300$ d.   

Interacting systems such as $\upsilon$ Andromedae are our most
powerful probes of the dynamical histories of exoplanets.  To select
just one example, \cite{Ford05} showed how $\upsilon$ Andromedae shows
good evidence that a single, strong, planet-planet scattering event is
the origin of the modest eccentricities of the outer two planets.

\subsubsection{HD 37124}

\cite{Vogt00}, using velocities obtained with HIRES at Keck Observatory, detected an
apparently Jupiter-mass planet with $P \sim 150$ d orbiting HD 37124,
a metal-poor G4 dwarf.  Further data showed significant deviations
from the predicted velocities, and allowed \cite{Butler03} to attempt
a double-planet fit including an outer planet with $P \sim 6$ y.
\cite{Gozdziewski03a} showed that this fit was unstable, and further
Keck data revealed the reason:  \cite{Vogt05} reported the detection
of a third planet in the system, though with an ambiguity in its
period.  Today, Keck data has resolved this ambiguity.

This system has been particularly difficult to unravel because the
radial velocity amplitudes of the three planets are similar and their
periods are long:  they
have $P \sim 155$, 840, and 2300 d, and $\msini =$ 0.64, 0.62, and
$0.68 \mjup$.  \cite{Gozdziewski06} find that the outermost planet's
orbit is still not very well constrained, and note that in this
system, planet-planet interactions are sufficiently strong that
kinematic (sum-of-Keplerian) models of the radial velocities are
not sufficient to describe the system.

\subsubsection{HD 74156}

\cite{Naef04} announced a double system with planets of $\msini =
1.86$ and $6.2$ \mjup\ in 52 d and 5.5 y orbits respectively around
HD 74156 from data taken with ELODIE at Observatoire de Haute
Provence. Combining published data from ELODIE, CORALIE at La Silla
Observatory, and new data acquired by HRS on HET, \cite{Bean08}
announced a third, intermediate planet with $\msini = 
0.4 \mjup$ and a period near 1 year.  \cite{Barnes08} studied the
system dynamically and found two, stable, qualitatively different
solutions to the published RV data.  They described the detection of
the third planet as vindication of a prediction by
\cite{Raymond05} of a planet of that mass and orbital distance based
on their ``Packed Planetary System'' hypothesis.

\subsubsection{$\mathbf \mu$ Arae = HD 160691}

\cite{Butler01} announced a $P \sim 700$ d planet with 2 \mjup\ orbiting $\mu$
Ara based on radial velocities from UCLES on the Anglo-Australian
Telescope, and soon thereafter \cite{Jones02} announced that further
observations revealed a linear trend in the residuals, indicative of a
long-period, outer companion.  Further AAT observations allowed
\cite{McCarthy04} to update the fit, note that the linear trend
then showed clear signs of curvature, and suggest a family of possible
orbits including one with $P =$ 8.2 y and $\msini = 3.1 \mjup$.  

Nearly simultaneously to the latter work, \cite{Santos04}, using HARPS
at La Silla Observatory to perform high-cadence, high-precision radial
velocity work, detected a $\msini = 10 \mearth$ companion in an inner
9.6 d orbit, and confirmed the outer planet.  Finally, \cite{Pepe07}
used a combination of old and new HARPS data and the published UCLES
data to detect a fourth planet with P=310 d, $\msini = 0.5 \mjup$ and
determine good orbital parameters for the outer planet for the first
time in a full, dynamical, 4-planet fit.\footnote{\cite{Gozdziewski07}
  also announced a tentative detection of the 310-d planet from a
  reanalysis of AAT data nearly simultaneously to \cite{Pepe07}.}  For the outer planet they
found $\msini = 1.8 \mjup$ and $P =$ 11.5 y, and they revised the orbit of the
$b$ component, finding $\msini = 1.7 \mjup$ and $P =$ 643 d.

Note that there is ambiguity in the literature regarding the
nomenclature for these planets, with 
some authors referring to the outer planet as the $e$ component (since
it was the last to be characterized), and others referring to it as
the $c$ component (since it was the second to be detected).  

\subsubsection{55 Cancri = $\rho^1$ Cancri}

Along with $\upsilon$ And $b$, \cite{Butler97} also announced 55 Cnc $b$, a
``51 Pegasi-type'' planet with $\msini = 0.84 \mjup$ and $P =$ 14.6 d.
Further Lick data allowed \cite{Marcy02} to announce a second planet
at 5 AU with $\msini = 4 \mjup$, the first extrasolar Jupiter analog (in terms of
orbital distance).  They also announced a signal from what appeared to
be a third planet with $P \sim 45$ d (roughly three times the period of
the inner planet) and $\msini \sim 0.2 \mjup$, but at the time stellar
rotation could not be ruled out as the cause. 

Using a combination of new and old Lick, ELODIE, and HET velocities
and Hubble Space Telescope Fine Guidance Sensor astrometry,
\cite{McArthur04} confirmed the 42-d planet and announced a very low
amplitude 2.8-d planet with $\msini = 14 \mearth$, one of the first of
a new class of ``Hot Neptunes''.  \cite{Wisdom05}, analyzing the
published radial velocity data, challenged the reality of the 2.8 d
planet, and noted a weak 260-d signal possibly due to a planet with
$\msini \sim 30 \mearth$.

Using a combination of new and old Lick and Keck data,
\cite{Fischer08} found a good orbital solution for all four published
planets and announced the fifth, $\msini = 0.14 \mjup$, 260-d planet,
making 55 Cnc the first (and, to date, only) known quintuple planet system.
They also found that despite the near-commensurability of their orbital
periods, the $b$ and $c$ components are not likely in a 3:1
mean-motion resonance because a dynamical integration shows that their
resonant arguments do not librate.  They note that, as with HD 37124,
Keplerian models are inadequate descriptions of the existing RV data,
and differ from the best Newtonian (dynamical) fits by $>$ 25 m/s.

\subsection{Resonant Doubles}

These six systems contain two giant planets in or suggested to be in
mean-motion resonances (MMRs).  While it is difficult to understand
how such planets could have formed {\it in situ}, differential
migration could explain how planets formed outside of resonance could
become trapped in such an MMR (e.g. \cite{Lee01}).  The frequency and
character of such MMR systems could thus be a probe of the nature of
planetary migration in systems with multiple giant planets.  

\subsubsection{GJ 876}

\cite{Marcy98}, using data from Keck Observatory, and
\cite{Delfosse98}, using ELODIE and CORALIE data, nearly
simultaneously announced the presence of a 61-d, 
$\msini \sim 2 \mjup$ planet orbiting GJ 876, the first known M-dwarf
planet host.  After 2.5 y of further observations at Keck,
\cite{Marcy01} showed that the signal was actually the superposition of
signals from two planets in a 2:1 mean-motion resonance, with the
inner planet having P= 30 d and $\msini = 0.6 \mjup$, making GJ 876
$b$ and $c$ the first system clearly shown to be in a mean motion
resonance.  Further Keck data allowed \cite{Butler04} to detect
a very low-amplitude, low-mass planet in a 2.6-d orbit.

This MMR is so strong that the orbital elements of the planets change
on timescales shorter than the span of the extant observations of the
system.  The arguments of periastron of the two components precess in
$\sim 11$ y, an effect clearly seen in the radial velocities and which
complicates multi-component Keplerian fits.  A dynamical fit of the Keck velocities
based on numerical integrations allowed \cite{Rivera05} to
weakly constrain the inclination of the system and estimate the true mass
of the inner planet to be $7.5 \mearth$.

\subsubsection{HD 82943}

\cite{Mayor04} described their discovery of the second known pair of
planets in a 2:1 MMR orbiting HD 82943 (which had been publicly
announced in 2000 and 2001) based on data from CORALIE.  \cite{Lee06} 
combined the published CORALIE data with new Keck data to derive a
dynamical solution to the system, showing that the only stable
solutions consistent with the data are those describing a 2:1 MMR.
The planets have $P =$ 219 and 441 d and $\msini = $ 2.0 and 1.8 \mjup, respectively.

\subsubsection{HD 128311}

From an analysis of Keck data, \cite{Butler03} announced a $P =$ 422 d, $\msini = 2.2 \mjup$ planet
around the chromospherically active K dwarf HD 128311, for which high
precision ($< 10$ m/s) can be difficult due to stellar ``jitter''.
\cite{Vogt05} used additional Keck data to detect an outer companion
of similar amplitude with $P = 928$ d and $\msini = 3.2 \mjup$, and used
dynamical simulations to show that the system is almost certainly
locked in a 2:1 MMR.  \cite{Sandor06} suggested that since the system
appears not to show apsidal corotation, it may owe its present state
to a strong scattering event in its past.

\subsubsection{HD 73526}

\cite{Tinney03} used data from UCLES to report the detection of a $P =
188$ d planet orbiting HD 73526.  Another three years of data allowed
\cite{Tinney06} to report a second planet in a 378 d orbit and show
that these planets ($\msini = 2.9$ and $2.5 \mjup$) are in 2:1
resonance.  \cite{Sandor06} showed that the published solution was
chaotic found alternative, non-chaotic (regular) orbital solutions
for the system, and argued that, like HD 128311, the system's dynamical state showed
evidence of a perturbative event such as a strong scattering event.

\subsubsection{HD 108874: 4:1 MMR?}

\cite{Butler03} used Keck data to report a $P \sim 400$ d Jovian
planet orbiting HD 108874, and noted that a good fit required a linear
trend be used in the model, suggesting a more distant companion was
present in the system.  By mid-2005, these residuals had turned over,
revealing a $P = 1600$ d outer companion.  They noted that while these
orbital periods are consistent with a 4:1 MMR, that resonance is
narrow and non-resonant stable configurations consistent with the data
exist. 

\subsubsection{HD 202206: 5:1 MMR?}

\cite{Udry02} used CORALIE data to detect a large ``superplanet'' or
brown dwarf with $\msini = 17.5 \mjup$ orbiting HD 202206 in a 255-d
orbit, one of only a few detections in the ``brown dwarf desert''.
Further CORALIE data allowed \cite{Correia05} to announce a second,
outer companion with $P = 3.8$ y and $\msini = 2.4 \mjup$.  They
report that while the system experiences strong planet-planet
interactions, stability is protected by a 5:1 resonance.

This system is especially interesting because it resembles a
circumbinary planet.  \cite{Gozdziewski06} confirmed the 5:1 MMR and
showed that the system is dynamically quite interesting, having
different qualitative behavior depending on the inclinations of the companions.  

\subsection{Interacting Doubles}

These eight systems contain two giant planets probably not in a {\em
  true} mean motion resonance, but for which other planet-planet
interactions can be important in the dynamical modeling of the system.
For instance, the planets orbiting HD 12661 are in apsidal libration. 


\subsubsection{HD 12661}

\cite{Fischer01} reported a planet orbiting the G6 dwarf HD 12661
based on observations at Lick and Keck Observatories.  Subsequent
observations allowed \cite{Fischer03} to report that the system is
actually a double.  Today's best parameters (\cite{Butler06}) show the
$b$ and $c$ components having $P = 263$ and 1822 d, and $\msini = 2.3$
and 1.8 $\mjup$, respectively. 

\cite{Gozdziewski03} found the system to be near the 6:1 MMR,
and referred to its ``Janus head'' of librating, anti-aligned apsidal lines.

\subsubsection{HD 155385}

\cite{Cochran07} used HET data to announce the lowest-metallicty
planet host, the G subgiant HD 155385.  The system is a double, with
no MMR but significant planet-planet interactions leading to
eccentricity exchange with a period of $\sim 2700$ y.  The $b$ and $c$
components have $P = 195$ and 530 d, and $\msini = 0.98$ and 0.5
$\mjup$, respectively.

\subsubsection{HD 169830}

\cite{Naef01} used CORALIE data to announce a $P \sim 230$ d planet
with $\msini = 2.9 \mjup$ orbiting the F8 dwarf HD 169830\footnote{Not
to be confused with HD 69830, the triple ``Super-Earth'' system of
\S\ref{69830}}.  Further data allowed \cite{Mayor04} to identify a
second, $P=5.8$ y, $\msini = 4 \mjup$ planet in the system.
\cite{Gozdziewski04} showed that there are large exchanges of
eccentricity between the components in coplanar configurations and
that the system is stable for a large range of inclinations.

\subsubsection{HD 183263}

\cite{Marcy05} used Keck data to report an $\msini = 3.7 \mjup$ planet
in a 634 d, eccentric orbit around HD 183263, and note a strong
residual linear trend.  \cite{Wright07} showed that by 2007 the residuals had
significant curvature, and \cite{Wright09} showed that the orbit,
though still incomplete, was sufficient, in combination with stability
analysis, to constrain the mass and period of the outer companion to
$P = 8.4 \pm 0.3$ y and $\msini = 3-4 \mjup$ under the assumption
that no additional companions are contributing to the radial velocities.

\subsubsection{47 Ursa Majoris}

The third star known to host exoplanets was 47 UMa, announced by
\cite{Butler96} as hosting a $P=3$ y planet on a circular orbit as
determined from data taken at Lick Observatory.  This
was the first planet strongly reminiscent of the gas giants in our
Solar System.  \cite{Fischer02} studied another 5 years of Lick data
and reported a second planet in the system, having $P \sim 1100$ d and
$\msini = 2.5 \mjup$.  \cite{Naef04} and \cite{Wittenmyer07}, using ELODIE and HET data,
respectively, have noted that a second planet with the reported
parameters is not apparent in their data.  Future observations should
clarify the situation.

\subsubsection{HIP 14810}

\cite{Butler06} included HIP 14810 $b$ in the {\it Catalog of Nearby
  Exoplanets} based on preliminary data from Keck Observatory taken as
part of the N2K survey (\cite{Fischer05}), and further observations allowed \cite{Wright07}
to provide an orbital solution for two planets, having $P = 6.7$ and
95 d and $\msini = 3.9$ and 0.76 $\mjup$.  Because the outer planet's
orbit is currently poorly sampled, further observations will help clarify the
nature of this system and allow for it to be better studied dynamically.

\subsubsection{OGLE-2006-BLG-109L}

\cite{Gaudi08} announced the remarkable detection of a double-planet
system around the distant ($d=1.5$kpc) star OGLE-2006-BLG-109L during a
microlensing event.  Due to the great distance to this system and the 
non-repeating nature of microlensing detections, the study of the
dynamics of this system with specificity is difficult (but see
\cite{Malhotra08}).  This detection demonstrates the promise of
microlensing as a method to build up statistics of multi-planet
systems, including those composed of rocky planets at a few AU.  The most likely
masses and orbital distances of these planets are $m \sim
0.71$ and $0.27 \mjup$ and $a\sim 2.3$ and 4.6 AU, making this system
around a K star a ``scaled-down Juipter-Saturn analog''

\subsubsection{HD 102272}

\cite{Niedzielski08} used HET to announce two companions orbiting the
giant star HD 102272.  The inner, $\msini = 5.9 \mjup$ planet orbiting
at 0.6 AU is the closest known companion to a star with $M > 1.5
M_\odot$.  Correlated residuals to a one-planet fit indicate a second
planet of uncertain orbital period and mass, but the velocities are
consistent with a period of $P \sim 520 $d.

\subsection{Highly Hierarchical Doubles}

These seven systems have orbital period ratios greater than 30:1, and
so have little interaction between their components and can usually be
well-modeled without resort to $N$-body simulations.

\subsubsection{HD 168443}

From Keck data, \cite{Marcy01b} discovered a pair of massive, highly
hierarchical companions 
orbiting HD 168443:  a close-in $P=58$ d inner planet with $\msini =
8.16 \mjup$ and an outer object with $P = 4.8$ y and $\msini
= 18.4 \mjup$.  Like HD 202206, this system has wide separation
between its components and contains a ``super-planet'', but here the
orbit of the lighter object is of the``S-type'' (it
orbits one, not both of the massive companions, \cite{Dvorak83}).

\subsubsection{HD 187123}

Using Keck data, \cite{Butler98} announced an $\msini = 0.5\mjup$ planet in a 3-day
orbit around HD 187123, a solar ``twin''.  Further observations
allowed \cite{Wright07} to report the existence of an outer companion
with orbital period $\geq 10$ y, and to constrain its minimum mass to
be planetary.  Using subsequent observations of the apparent closing
of the orbit, \cite{Wright09} constrained its orbit to have $P = 10.5
\pm 0.5$ y and $\msini = 2.0 \pm 0.1 \mjup$, under the assumption that
no other planets are influencing the observations.

\subsubsection{HD 68988}

\cite{Vogt02} used Keck data to report a 6.3-d, $\msini=2 \mjup$
planet orbiting HD 68988.  Subsequent observations have revealed a
long-period outer companion of uncertain mass and period, with
\cite{Wright07} constrained to be $11 $y $< P < 60$ y and $6 \mjup <
\msini < 20 \mjup$.

\subsubsection{HD 38529}
\cite{Fischer01} used Lick and Keck data to report a 14.3-d, $\msini =
0.8 \mjup$ companion to the old G subgiant HD 38529, and noted that the residuals to the
fit suggested an outer companion.  Further observations allowed
\cite{Fischer03} to confirm an outer, $P=5.9$ y, $\msini = 13.2 \mjup$
companion.  Spitzer observations by \cite{MoroMartin07} reveal that HD
38529 has infrared excess consistent with dust-producing
planetessimals at around 5 AU (exterior to both planets).

\subsubsection{HD 217107}

\cite{Fischer99} used Lick data to discover the first, inner $P=7.1$
d, $\msini = 1.3 \mjup$ planet around the G7 dwarf HD 217107.
\cite{Vogt05} used new and old Lick and Keck data to identify a long
period outer companion of uncertain mass and period.  Further
observations at Keck allowed \cite{Wright09} to constrain the outer
companion's mass and period to be $P \sim 11.7$ y and $\msini \sim
2.6 \mjup$ under the assumption that no other planets are contributing
to the radial velocities.

\subsubsection{HD 11964}

\cite{Butler06} announced a $\msini = 0.6 \mjup$ planet orbiting the
slightly evolved star HD 11964 in a 5.3 y orbit.  Subsequent
observations suggested a low mass inner planet with $P= 38$ d (\cite{Wright07}), and
further monitoring allowed \cite{Wright09} to confirm this signal as
being to due a $\msini = 23 \mearth$ companion.

\subsubsection{GJ 777 A = HD 190360}

\cite{Naef03} used data from ELODIE and the AFOE spectrograph at
Whipple Observatory to detect\footnote{A preliminary orbit for this
  system also appears in \cite{Udry03}.} a long-period, Jovian planet around GJ
777 A.  \cite{Vogt05} used Keck data to confirm the outer planet and
revise its orbital parameters, and to announce a second, close-in,
low-mass companion.  The $b$ and $c$ components have $P = 8.0$ y and
17.1 d, and $\msini = 1.6 \mjup$ and $19 \mearth$.

\subsection{Systems of ``Rocky'' Planets and ``Super-Earths''}

These four systems compose the present-day bookends of
multiple-planet systems.  The pulsar triple-planet system is the first
exoplanetary system known and remains a fascinating example of the
limits of multiple-planet detection.  The latest frontier of 
exoplanet research with radial velocities is the hunt for rocky
planets, and these latest detections of multiple ``Super-Earths'', all
from the HARPS spectrograph, represent the penultimate step toward the
definitive detection of rocky planets.

Because they do not transit, the actual compositions and masses of
these planets is unknown, and so the monicker ``rocky'' is probably
only truly appropriate for the pulsar planets.  

\subsubsection{PSR B1257+12}

\cite{Wolszczan92} detected two very low-mass ($\msini = 3.9$ and 4.3
\mearth) objects orbiting the pulsar PSR B1257+12 using pulse timing
methods.  Two years later, \cite{Wolszczan94} detected the
planet-planet interactions from the planets' 3:2 MMR, as well as a
smaller, 0.02 \mearth\ object in the system.  

\subsubsection{HD 69830}
\label{69830}

Based on observations with HARPS, \cite{Lovis06} announced a
``triple-Neptune'' system orbiting the K dwarf HD 69830, having
periods of $\sim$ 9, 32, and 200 days.  This system is especially
interesting because \cite{Beichman05} had used Spitzer photometry and
spectroscopy to reveal an infrared excess characteristic of a large
cloud of fine silicate dust within a few AU of the star, suggestive of
a large asteroid belt or ``super-comet''.  

\subsubsection{GJ 581}

Also using HARPS, \cite{Bonfils05} announced a $\msini = 0.052 \mjup$
planet in a 5.366 d orbit around GJ 581, making that star only the
third M dwarf known to host a planet.  After collecting additional data, two
years later \cite{Udry07} announced the discovery of two additional
planets in the system with minimum masses of 5 and 7.7 \mearth, with
periods of 12.9 and 83.6 d.  The outermost planet in the system sits
at 0.25 AU, a location \cite{Udry07} and \cite{Selsis07} identify as
being near the ``cold edge'' of the star's Habitable Zone
(\cite{Kasting93}).

\subsubsection{HD 40307}

\cite{Mayor08} announced a third triple ``Super-Earth'' system from
HARPS around the K2, dwarf HD 40307, and noted
a linear trend in the radial velocities suggestive of a fourth, outer
companion, as well.  The low metallicity of this star ([Fe/H]
$=-0.31$) and others has led to the suggestion that the well known
metallicity dependence of planet occurrence breaks down for low-mass stars.

\subsection{Future Detections and the Multiplicty Rate}

Many systems show strong evidence of second planets due to long-period
companions whose orbits are too incomplete for strong constraints to
be put on their masses.  For instance, 14 Her clearly has a long
period companion of some sort, probably planetary (\cite{Naef04},
\cite{Wittenmyer07}, \cite{Wright07}, and \cite{Gozdziewski08}), as
may GJ 317 (\cite{Johnson07}).  

Of the 200 planet-bearing stars within 200 pc, the 28 above (not
counting PSR B1257+12 and OGLE-2006-BLG-109L) constitute 14\% of the
total.  An additional 27 (including 14 Her and GJ 317) show clear
evidence of trends in their residuals and no evidence of stellar
duplicity, meaning that the true multiplicity rate is at least 28\%.

\section{Statistical Properties of the Multiple Planet Sample}

\begin{figure} [ht!]
\begin{center}
  \includegraphics*[width=5in]{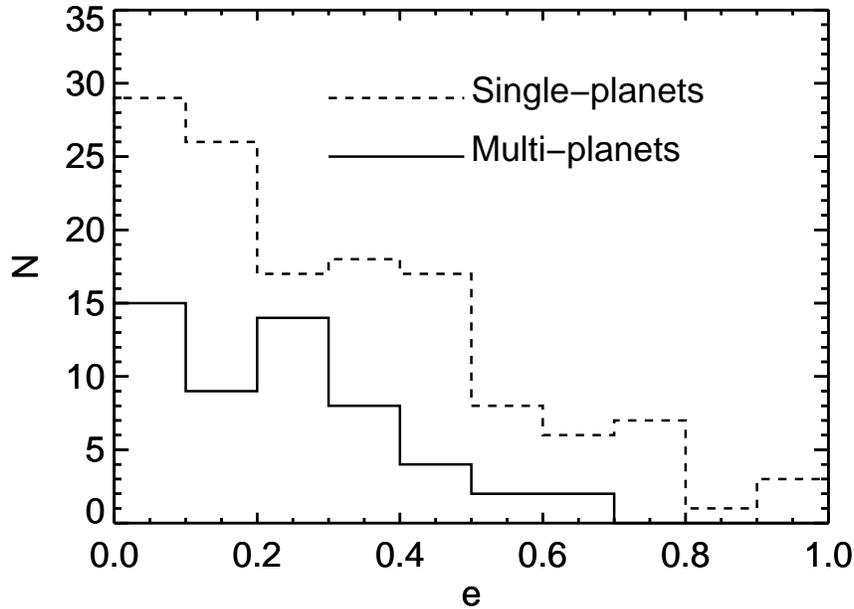}
  \caption{Distribution of eccentricities of exoplanets for known multiple
    planet systems (solid) and apparently single planet systems
    (dashed).  Note the high eccentricity orbits, $e>0.6$ occur
    predominantly in single planets.}\label{e_hist}
\end{center}
\end{figure}

\begin{figure} [ht!]
\begin{center}
\includegraphics*[width=5in]{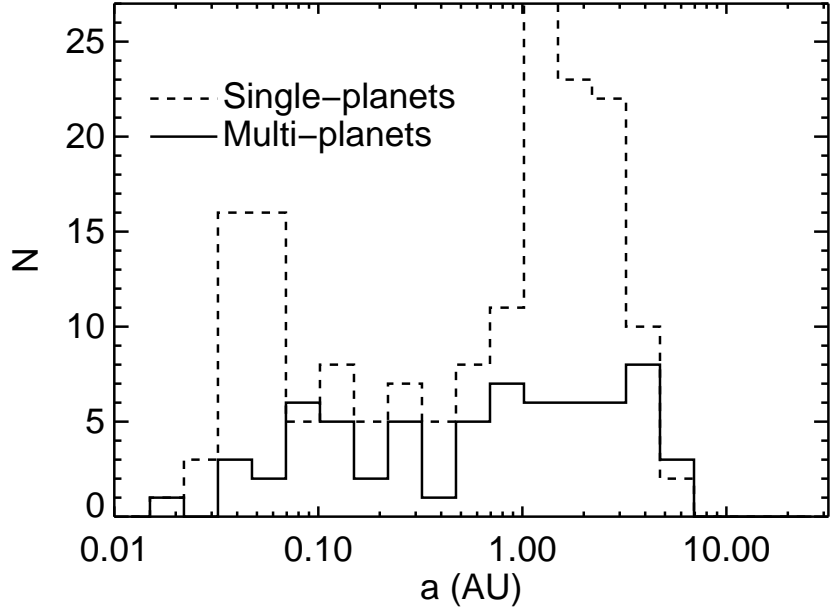}
  \caption{Distribution of semimajor axes of exoplanets for
    multiple-planet systems (solid) and apparently single systems
    (dashed).  Note the enhanced frequency of hot jupiters and the
    jump in abundance beyond 1 AU in the single-planet systems.}\label{a_hist_log}
\end{center}
\end{figure}

\cite{Wright09} contains a catalog of the above multi-planet systems
including updated latest orbital parameters for 10 of the systems.
From this catalog, several intriguing patterns emerge;  these features
represent an opportunity to test models of planet formation,
migration, and the origin of eccentricities:    
\begin{itemize}
\item Planets in multiple-planet systems have eccentricities no higher
  than single planets (see Figure~\ref{e_hist}).   
\item The distribution of orbital distances of planets in multi-planet systems
and single planets are inconsistent:  single-planet systems show a
pile-up at $P \sim 3$ days and a jump near 1 AU, while multi-planet
systems show a more uniform distribution in log-period (see
Figure~\ref{a_hist_log}).   
\end{itemize}

\begin{figure} [ht!]
\begin{center}
  \includegraphics*[width=5in]{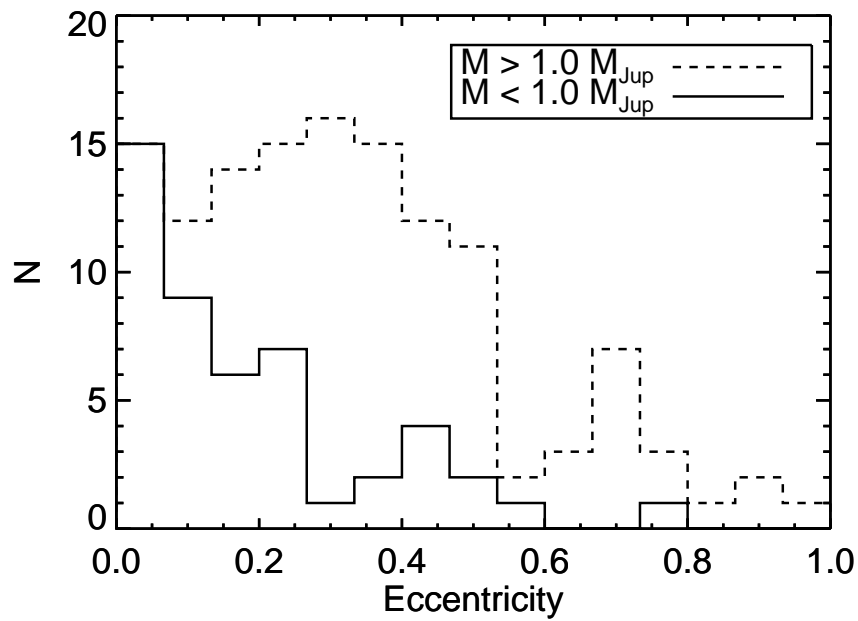}
  \caption{Distribution of eccentricities of exoplanets with $\msini <
    1.0 \mjup$ (solid) $\msini > 1.0 \mjup$ (dashed).  The tidally
    circularized hot jupiters have been removed. Note that the
    eccentricity of planets of minimum mass $ < 1.0 \mjup$ peaks at
    eccentricity $< 0.2$, while the eccentricities $e$ of planets of
    minimum mass $ > 1.0 \mjup$ are distributed broadly from
    $0.0<e<0.5$.}\label{eccmass}
\end{center}
\end{figure}

In addition, among all planetary systems:
\begin{itemize}
\item There may be an emerging, positive correlation between stellar
  mass and giant-planet semi-major axis (see, e.g., \cite{Johnson07}).   
\item Exoplanets more massive than Jupiter have eccentricities
broadly distributed across $0 < e < 0.5$, while lower-mass exoplanets
exhibit a distribution peaked near $e=0$ (see Figure~\ref{eccmass}).
\end{itemize}

From Figure~\ref{a_hist_log} it is clear that the orbital distances of
planets in systems currently known to be multiple are not drawn from
the apparently single-planet distribution, indicating that the
migration mechanisms operate differently in these populations.  Yet
the eccentricity distributions of these two populations are nearly
identical (Figure~\ref{e_hist}), suggesting that the mechanisms of
eccentricity excitation are similar.  Figure~\ref{eccmass} shows that
whatever those mechanisms are, they are ultimately more effective in
pumping the eccentricities of planets with $\msini > 1 \mjup$.
Reproducing these distributions will be a good test for future models
and simulations of planet formation and dynamical evolution.


\end{document}